\documentclass[12pt,twoside]{article}
\usepackage{fleqn,espcrc1,epsfig,citesort}

\usepackage{graphicx}

\newcommand{\gsim}{\mathrel{\mathop{\kern 0pt \rlap
  {\raise.2ex\hbox{$>$}}}
  \lower.9ex\hbox{\kern-.190em $\sim$}}}

\title{Mass--dependence of the $\Lambda$ hypernuclear decay widths}
\author{G. Garbarino \address{Grup de F\'{\i}sica Te\`{o}rica,  
        Universitat Aut\`{o}noma de Barcelona, \\ 
        08193 Bellaterra, Barcelona, Spain}\thanks{The author acknowledges 
        financial support from the EEC through TMR Contract CEE--0169}}

\begin{document}
\maketitle

\begin{abstract}
Two different approaches have been employed for the evaluation of the decay 
widths of $\Lambda$--hypernuclei (ranging from $^5_{\Lambda}$He to 
$^{208}_{\Lambda}$Pb) with the polarization propagator
method. In ref.~\cite{Ga1}, the two--nucleon
stimulated non--mesonic decay, $\Lambda NN\to NNN$, has been parameterized
phenomenologically
by means of data on the pion--nucleus optical potential. The other approach
\cite{Ga2} consisted in a fully microscopic description of the 
non--mesonic decays through the first order approximation
of the so--called bosonic--loop--expansion. Both calculations reproduce,
with approximately the same accuracy, the experimental decay rates 
for the whole range of mass numbers considered. 
\end{abstract}

\section{INTRODUCTION}

Because of its kinematics, the decay mode of the free $\Lambda$--hyperon, 
$\Lambda \to \pi N$, is disfavoured by the Pauli principle if it
occurs in nuclear systems. However, the presence of the nuclear medium
is responsible for the opening of new, non--mesonic decay channels,
$\Lambda N\to NN$ and $\Lambda NN\to NNN$. The non--mesonic decay 
is characterised by large momentum transfers, thus the details of
nuclear structure do not have a substantial influence, but the
$NN$ and $\Lambda N$ short range correlations turn out to be very important.
Experimental data show an anticorrelation between mesonic and non--mesonic
decay modes such that the total lifetime is quite stable over the 
whole hypernuclear mass spectrum. 

\section{FRAMEWORK FOR CALCULATION}

Within the polarization propagator method the weak decay 
of $\Lambda$--hypernuclei is studied through
a many--body description of the hyperon self--energy, 
whose imaginary part gives the $\Lambda$ decay width:
\begin{equation}
{\Gamma}_{\Lambda}=-2\;{\rm Im}\,{\Sigma}_{\Lambda} .
\end{equation}
This technique provides a unified picture of the different decay
channels. Moreover, it is alternative and equivalent to the
standard wave function method, which makes use of shell model
hypernuclear wave functions and pion waves generated by
pion--nucleus optical potentials. 

Here we only present and discuss
the results of our calculations. Details concerning the framework
used can be found in refs.~\cite{Ga1,Ga2,Os85}. The $2p-2h$ part of
the irreducible polarization propagator has been estimated, in 
ref.~\cite{Ga1}, by using a {\it phenomenological} parameterization supplied 
by the pion--nucleus optical potential, extracted from data on pion absorption
in pionic atoms. The phase space available for real $2p-2h$ excitations
has also been taken into account, in order to extrapolate the data for
off--mass shell pions. The Feynman diagrams contributing to the $\Lambda$
self--energy in nuclear systems can be classified in a theoretically
grounded scheme by using a functional approach, according to the prescriptions
of the so--called bosonic--loop--expansion. In ref.~\cite{Ga2},
a {\it microscopic} calculation of the non--mesonic decays within the
one--boson--loop approximation has been performed.

\section{RESULTS}

In this section the main results obtained in refs.~\cite{Ga1,Ga2} 
are discussed. 

\subsection{Phenomenological calculation}

The local density approximation, used to extend the calculation
made in nuclear matter to finite nuclei, requires the knowledge
of the $\Lambda$ wave function in nuclei. We have obtained 
these wave functions from Wood--Saxon wells with radius and depth
fixed to reproduce the first two single particle states of the
hypernuclei considered.

A crucial ingredient in the calculation of the non--mesonic decay rates
is the short range part of the strong $NN$ and $\Lambda N$ interactions.
They have been expressed by phenomenological functions, which
supply a good reproduction of realistic $G$--matrix calculations
in the $NN$ case. Since 1) there are no experimental indications on 
the strong $\Lambda N$ short range
interaction and 2) we use a phenomenological picture of the $\Lambda$
decay which go beyond the description usually employed
for processes not involving strangeness, the zero energy and momentum limit,
$g'$ and $g'_{\Lambda}$,
of the correlation functions have been kept as free parameters. 
They have been fixed (on the values $g'=0.8$ and $g_{\Lambda}'=0.4$)
in order to reproduce the non--mesonic
decay rate of $^{12}_{\Lambda}$C. 

\begin{figure}[thb]
\begin{center}
\mbox{\epsfig{file=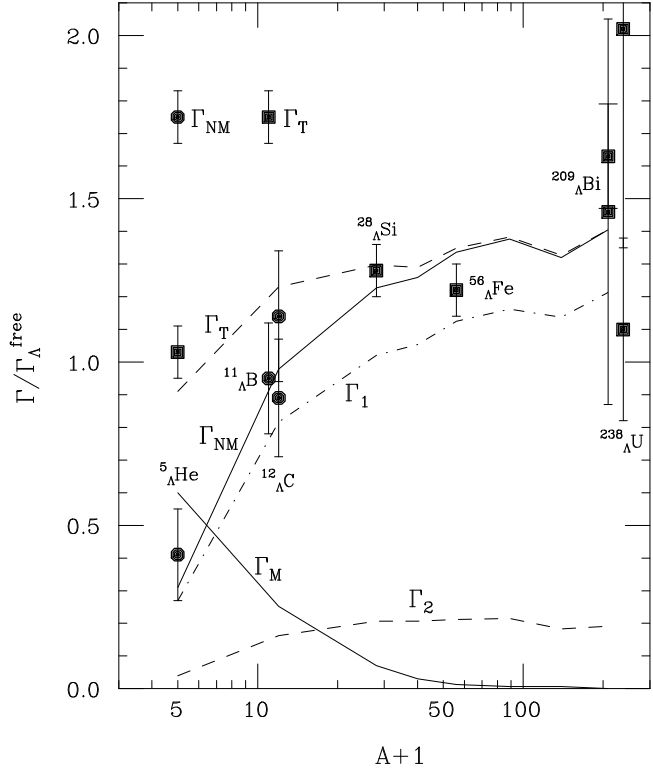,width=.55\textwidth}}
\vskip -9mm
\caption{Partial ${\Lambda}$ decay widths in finite nuclei as a function of
the mass number $A$.}
\vskip -9mm
\label{satu}
\end{center}
\end{figure}

Then, the calculation has been extended 
to hypernuclei ranging from $^5_{\Lambda}$He to $^{208}_{\Lambda}$Pb.
The result are shown in figure \ref{satu}.
The mesonic rate $\Gamma_M$ rapidly 
vanishes with the nuclear mass number $A$. This
is related to the decreasing phase space allowed for the decay and
to smaller overlaps between the $\Lambda$ probability distribution 
and the nuclear surface, as $A$ increases.
The results for $\Gamma_M$ are in good agreement both with experimental data 
(included the new ones from KEK presented at this conference \cite{Sa00}) 
and with other
theoretical estimations \cite{Os93,Mo94}, obtained in shell  model
calculations. With the exception of $^5_{\Lambda}$He, the two--body induced
decay ($\Gamma_2$) is rather independent of the hypernuclear dimension
and is about 15\% of the total width, which is also fairly constant
with $A$. In figure \ref{satu} our calculation for non--mesonic and total
decay widths ($\Gamma_{NM}=\Gamma_1+\Gamma_2$ and $\Gamma_T$) is compared with
recent (after 1990) experimental data \cite{Sz91,Ar93,No95,Ku98,Pa00}. 
The theoretical results are in good agreement with the experiment 
over the whole hypernuclear mass range explored. In particular,
the saturation of the $\Lambda N\to NN$ interaction in nuclei is well
reproduced.

\subsection{Microscopic calculation}

The results presented in this section have been obtained in 
ref.~\cite{Ga2} within the one--boson--loop (OBL) approximation.
The calculation of the OBL diagrams for the $\Lambda$ self--energy
in finite nuclei (using the local density approximation) 
is not possible here because of the long computing time
already for the evaluation of the decay rates 
at fixed Fermi momentum (namely in nuclear matter).
In order to compare the calculation with experimental
data, we assigned 
different ``average'' Fermi momenta to three mass regions in the hypernuclear
spectrum. We used the following prescription:
\begin{equation}
\label{kf3}
\langle k_F\rangle=\int d\vec r \,k_F(\vec r)|\psi_{\Lambda}(\vec r)|^2 ,
\end{equation}
which derives from weighting the nucleon local Fermi momentum $k_F(\vec r)$
with the probability density of the $\Lambda$ inside the nucleus
$|\psi_{\Lambda}(\vec r)|^2$.
We classified the hypernuclei for which experimental data are available
into the following mass regions:
\begin{itemize}
\item medium--light: $A\simeq 10$ $\Rightarrow$ $\langle k_F\rangle=1.1$ fm$^{-1}$;
\item medium: $A\simeq 30\div 60$ $\Rightarrow$ $\langle k_F\rangle=1.2$ fm$^{-1}$;
\item heavy: $A\gsim 200$ $\Rightarrow$ $\langle k_F\rangle=1.36$ fm$^{-1}$ .
\end{itemize}
As in the phenomenological approach, the Landau parameters 
$g'$ and $g_{\Lambda}'$ are the only free parameters of the
calculation. 

\begin{figure}[thb]
\begin{center}
\mbox{\epsfig{file=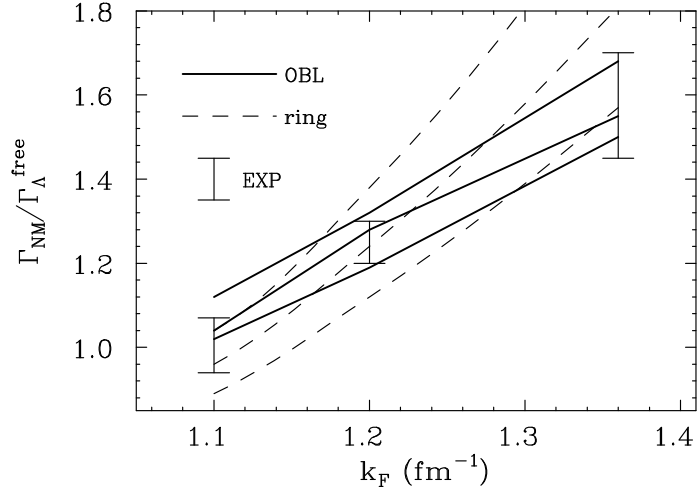,width=0.6\textwidth}}
\vskip -9mm
\caption{Dependence of the non--mesonic width on the Fermi momentum.
The solid curves refer to the one--boson--loop approximation
(with $g^{\prime}=0.7, 0.8, 0.9$ from the top to the bottom), 
while the dashed lines refer to the ring 
approximation ($g^{\prime}=0.5, 0.6, 0.7$). 
Reference experimental bands are also shown.}
\vskip -9mm
\label{micro}
\end{center}
\end{figure}

In figure \ref{micro} we show the dependence of the total non--mesonic 
width on the Fermi momentum of nuclear matter
(we remind that for this range of Fermi momenta,
in infinite nuclear matter
the mesonic decay is strictly forbidden). Here $g_{\Lambda}'$
has been fixed to 0.4, since with this value the experimental
data can be reproduced, in ring approximation,
by using $g'$ values compatible
with the phenomenology of other physical processes.
The solid curves refer to the OBL approximation,
with $g^{\prime}=0.7$, 0.8, 0.9 from the top to the bottom, 
while the dashed lines refer to the ring approximation, with
$g^{\prime}=0.5$, 0.6, 0.7, again from the top to the bottom. 
We note that the OBL calculation reproduce rather well the data
for the three mass regions when $g'=0.8$.
Incidentally, the same value for $g'$ has been employed in the
phenomenological calculation. However, one must point out that
the role played by the Landau parameters is different in the
two approaches we used.

In conclusion, both calculations reproduce, with approximately the
same accuracy, the experimental data. 
This proves the reliability
in using 1) averaged Fermi momenta to simulate the $\Lambda$ decay
in finite nuclei and 2) the phenomenological model to
describe the non--mesonic decay.


\begin{thebibliography}{9}
\bibitem{Ga1} W. M. Alberico, A. De Pace, G. Garbarino and A. Ramos,
Phys. Rev. {\bf C 61} (2000) 044314.
\bibitem{Ga2} W. M. Alberico, A. De Pace, G. Garbarino and R. Cenni,
Nucl. Phys. {\bf A 668} (2000) 113.
\bibitem{Os85} E. Oset and L. L. Salcedo, Nucl. Phys. {\bf A 443} (1985) 704.
\bibitem{Sa00} Y. Sato, these proceedings.
\bibitem{Os93} U. Straub, J. Nieves, A. Faessler and E. Oset,
Nucl. Phys. {\bf A 556} (1993) 531.
\bibitem{Mo94} T. Motoba and K. Itonaga, Prog. Theor. Phys. Suppl. {\bf 117}
(1994) 477.
\bibitem{Sz91}J. J. Szymanski {\sl et al.}, Phys. Rev. {\bf C 43} (1991) 849.
\bibitem{Ar93} T. A. Armstrong {\sl et al.}, Phys. Rev. {\bf C 47} (1993) 1957.
\bibitem{No95} H. Noumi {\sl et al.}, Phys. Rev. {\bf C 52} (1995) 2936.
\bibitem{Ku98} P. Kulessa {\sl et al.}, Phys. Lett. {\bf B 427} 
(1998) 403; H. Ohm {\sl et al.}, Phys. Rev. {\bf C 55} (1997) 3062.
\bibitem{Pa00} H. Park {\sl et al.}, Phys. Rev. {\bf C 61} (2000) 054004.

\end{thebibliography}
\end{document}